\documentstyle[preprint,aps]{revtex} 

\def\plb#1#2#3{ Phys. Lett. {\bf B #1} (#2) #3}
\def\npb#1#2#3{ Nucl. Phys. {\bf B #1} (#2) #3}
\def\prd#1#2#3{ Phys. Rev.  {\bf D #1} (#2) #3}
\def\prl#1#2#3{ Phys. Rev. Lett. {\bf #1} (#2) #3}
\def\zpc#1#2#3{ Z Phys. {\bf C #1} (#2) #3}
\def\nca#1#2#3{ Nuovo Cimento {\bf A #1} (#2) #3}

\def\alms#1{\mbox{$\widehat{\alpha}_#1^{-1}$}}
\def\alz#1{\mbox{$\alpha_#1(m_Z)^{-1}$}}
\def\alzi#1{\mbox{$\alpha_#1(m_Z)$}}
\def\alzms#1{\mbox{$\widehat{\alpha}_#1(m_Z)^{-1}$}}
\def\almsi#1{\mbox{$\widehat{\alpha}_#1$}}
\def\alzmsi#1{\mbox{$\widehat{\alpha}_#1(m_Z)$}}
\def\ms{\mbox{$\overline{MS}$}}
\def\mz{\mbox{$m_Z$}}
\def\mzz{\mbox{$m_Z^2$}}
\def\su5{\mbox{$SU(5)$}}
\def\mv{\mbox{$M_V$}}
\def\mx{\mbox{$M_X$}}
\def\mphi{\mbox{$M_{\Phi}$}}
\def\msig{\mbox{$M_{\Sigma}$}}
\def\mp{\mbox{$M_{P}$}}
\def\sf{{\scriptscriptstyle F}}
\def\sg{{\scriptscriptstyle G}}
\def\ss{{\scriptscriptstyle S}}

\def\sgam{{\scriptscriptstyle \Gamma}}
\begin{document}



\nopagebreak
\title
{
 Perturbative Unification and Low Energy Precision Data. \\
} 
\author{M. Bastero--Gil} 
\vspace{.125in}

\address{ Scuola Internazionale Superiore di Studi Avanzati \\ 34013
Trieste, ITALY. 
}
\vspace{.125in}


\maketitle


\begin{abstract} 
More precise unification predictions require going beyond the lowest
order, including 2--loop running of the couplings and a correct
treatment of threshold effects. Here we revised two different
approaches to deal with light thresholds, based on different choices
of the renormalization scheme, \ms\ and effective couplings. We show
the equivalence of both approaches in making predictions when
thresholds are taking properly into account. 
\end{abstract} 

\vspace{.125in}

\narrowtext
\section{Introduction}

 Experimental data have always played a relevant role in building
unification scenarios.  Whereas Standard $SU(5)$ unification
\cite{su5} was not  
completely ruled out by experiments in the early 80's, now
attempts to unify without introducing new degrees of freedom between
the electroweak scale and the unification scale does not
work. Introduction of  new degrees of freedom modifies $\beta$--functions,
and opens up the possibility for unification. As a general result
\cite{terry}, one is constrained to introduce new physics at a
intermediate scale $O(10^8-10^{12}\,GeV)$ \cite{iscale}, and/or to
populate the spectrum with many new degrees of freedom at the scale of
$O(1\,TeV)$. The second alternative is provided by the Minimal
Supersymmetric extension of the Standard Model (MSSM). Supersymmetric
Grand Unification Theories (Susy GUT) \cite{susygut} have been widely
studied in the literature, in both versions, with \cite{so10i} and without
\cite{so10,ssu5,hisano,gutsf} intermediate scale.

Due to the presence of a rich susy spectrum at low energies (with
masses no more than a few $TeV$), important light threshold
contributions come into the game.  The simplest procedure to deal with
them consists of using the step--function approximation (or
run--and--match procedure), so that a
particle contributes to the evolution of the couplings only beyond its
mass scale, giving zero contribution otherwise. However, the
uncertainty principle tells us that the effects from a particle are
felt not only beyond its excitation from the vacuum but, since the
packet necessarily spreads in momentum, its contribution to the
physical processes will occur even before its mass scale is reached,
contrary to the assumption of the step--function approach.  In order
to get a more accurate description of the threshold behavior, one can
compute the running couplings working with a Mass Dependent
Subtraction Procedure (MDSP) \cite{mdsp,clavelli}, or the equivalent effective
couplings \cite{effcoup,effqcd,mar},  
in which all the information about the mass spectrum is
automatically included in the $\beta$--functions.

Another approach related to the correct treatment of light susy
thresholds is that of Ref. \cite{grinstein}. They remark that the
extraction of the values of the gauge coupling constants at the \mz\
scale is not only renormalization scheme dependent but also {\it model
dependent}. The values extracted assuming the Standard Model (SM) will
not be equal to those extracted if we assume that the MSSM is correct, that is,
\begin{equation}
 \alpha_i(\mz)\mid_{MSSM} = \alpha_i(\mz)\mid_{SM} + \delta
\alpha_i(\mz)\mid_{\rm new ~physics}\,,
\label{eq1}
\end{equation}
where ``new physics'' means new degrees of freedom different from those
present in the SM. The question that arises again is to which
accuracy do we want to compute the function $\delta \alpha_i(\mz)$
\cite{bagger}, which includes those threshold effects due to 
the susy degrees of freedom. This function will also depend  on the
renormalization prescription used to define the gauge couplings.
Working with \ms\ but including $complete$ threshold effects in
$\delta\alpha_i$ we will get a non-zero contribution coming from new
physics, even for masses several orders of magnitude above \mz. With
the use of the \ms\ procedure, the decoupling theorem is not
implemented. The situation is different for the effective couplings, and
in this case those contributions coming from heavy degrees of freedom,
relative to \mz, are suppressed in $\delta\alpha_i$.

It is  clear by now that in studying susy unification, as far as all 
thresholds are crossed in going from \mz\ to $M_X$, different 
conclusions will be achieved when using an approximate treatment like  
the step--function \cite{gutsf}, or the more accurate one 
\cite{mar,bagger}. 
Moreover, in studying $complete$ threshold effects and the related
modification of the evolution of the gauge couplings, we have different 
approaches, depending on the {\it renormalization scheme} we 
choose. For example,

-- One can work with \ms, including the complete thresholds at the
electroweak scale throughout Eq. (\ref{eq1}). After that, we run the
couplings up to the high energy region just using the \ms\ $\beta$--
functions, without the need of any other consideration about the
mass spectrum.

-- Other choice is to work instead with effective couplings (equivalent
to work with a MDSP).  The values of 
the effective couplings at the electroweak scale \mz\ will be different
from those of the \ms\ couplings. At each scale $\mu \ge \mz$, the
contribution of a massive degree of freedom is controlled via a smooth
function $F(m/\mu)$ which gives an appropriate threshold crossing as we
evolve the couplings, and goes to zero for masses $m \gg \mu$.

As  physical quantities and physical conclusions must be
renormalization scheme independent, and in both approaches $complete$
thresholds are supposed to be included, the conclusions reached about
unification using one approach or another $should$ be the same.
In this work, we aim to show explicitly that in fact this is the case.
In particular, we will compare the prediction for the QCD coupling
\alz{3}\ in both schemes. We can not expect these values to be equal,
because they refer to different renormalization prescriptions. However,
we can calculate independently the relation between different
schemes. Therefore, we will recover the prediction of the effective
couplings approach from that of the \ms\ couplings and vice versa. 
 
This is the normal procedure to deal with physical processes and experimental
quantities. Using perturbation theory, they are written like a series
expansion in some parameter, say the coupling constant, in a given
renormalization scheme. Depending on the expansion parameter, the
coefficients in the series will be different, but the final result must
be the same (modulo higher order corrections). 

In Section 2 we will extract the values of the MSSM gauge couplings at
\mz\ in both schemes, effective couplings\footnote{Our work with
effective couplings is based in that of  
Ref. \cite{mar}. In that work, we took as initial data for the
effective couplings at \mz\ those values quoted by the Particle Data
Group \cite{pdg}, that is, those initial values valid for the SM when
using \ms. Here  we compute the correct 
initial values at \mz\ for the effective couplings.} 
and \ms. Comparing with the
values for the SM and \ms, we will see that the main source of the
differences is not due to susy threshold effects, but lies on the
renormalization prescription is used. The difference is more
pronounced for the value of the QCD coupling constant. 

In Section 3 we use these initial values to check the unification
scenario, and the equivalence of the predictions in both
approaches. When working with effective couplings, as we approach the 
high energy region, we would expect to feel the heavy degrees of
freedom coming from the unification group. These fields are needed in
order to get not only the same value, but the same evolution of the
couplings beyond some scale \mx. Their contributions depend on the
specific unification group considered. As far as we are not interested in
the study of a particular model, we will try to keep the discussion as
general as possible, but we will fix the unification group to be
$SU(5)$ when needed for numerical calculations. The inclusion of these
heavy threshold effects will also be relevant to get the same results
with effective couplings and \ms.

In section 4 we present our concluding remarks.

\section{Initial values at \mz}

The renormalized couplings using a MDSP are equivalent to effective
couplings, defined by \cite{mar}, 
\begin{eqnarray}
\alpha_i^{-1}(q^2)
&=&\alpha_{i0}^{-1}+\left( \Pi^T_i(q)+2 \Gamma_i^{U}(q) \right) \nonumber \\
&=&\alpha_i^{-1}(q_0^2) + \left(\Pi^T_{i}(q^2)+2
\Gamma^U_i(q^2) -\Pi^T_{i}(q_0^2)-2 \Gamma^U_i(q^2_0)\right)\,,
\label{eff}
\end{eqnarray}
where $\alpha_{i0}$ is the bare coupling constant, 
$\Pi^T_{i}$ is the transverse component of the bare vacuum
polarization tensor of the gauge boson defining the interaction, and
$\Gamma^U_i$ is the universal (process--independent) vertex
correction. All the dependence on the masses is included in these 
functions. The divergences of the bare function $\Pi_i^T$, and
$\Gamma^U_i$, cancel out in the differences, and we do not need any
additional assumption to render finite the above expression. 

The initial values\footnote{$\alpha_1^{-1}$ is given by the relation
$\alpha_1^{-1}=3 (\alpha_e^{-1}-\alpha_2^{-1})/5$, at any scale.}
$\alpha_e(\mz)^{-1}$ and $\alpha_2(\mz)^{-1}$ can be  
obtained from the set of experimental data $G_{\mu}=1.16639\times
10^{-5} \,GeV^2$, $\alpha_e(0)=1/137.036$ and $\mz=91.19\,GeV$. The
value $\alpha_e(\mz)^{-1}$ is obtained using $\alpha_e(0)$ and
Eq. (\ref{eff}) for $q_0^2=0$. That expression can be written in a
more familiar way like,
\begin{equation}
\alpha^{-1}_e(\mz)=\alpha^{-1}_e(0)\left[ 1+\Delta \alpha_e(\mz) \right]\,,
\end{equation}
where,
\begin{eqnarray}
\Delta \alpha_e(\mz)
  &=& \alpha_e(0) \left[ \Pi^T_{\gamma}(\mz)+2
\Gamma^U_{\gamma}(\mz) -\Pi^T_{\gamma}(0)-2
\Gamma^U_{\gamma}(0) \right] \nonumber \\
&=& \Delta \alpha_{leptons}(\mz)
+\Delta \alpha^{(5)}_{hadrons}(\mz)+\Delta \alpha_{top}(\mz)+\Delta
\alpha_{gauge}(\mz)+\Delta \alpha_{susy}(\mz)\,. 
\label{alphae}
\end{eqnarray}
All the terms in Eq. (\ref{alphae}) except that for the light quarks
$\Delta \alpha_{hadrons}^{(5)}$, can be computed in perturbation 
theory . For the latter,  we use the recent data \cite{had},
\begin{equation}
\Delta \alpha_{hadrons}^{(5)}(\mz)= 0.0280\pm 0.0007\,.
\end{equation}
 We also include the dominant fermionic 2-loops
contributions of $O(\alpha_e^2)$ and $O(\alpha_e \alpha_3)$ in the
other terms.
\cite{2loop}.  

 In order to obtain $\alpha_2(\mz)^{-1}$, we use the definition of
$G_{\mu}$ as the limit of the charged current process involved in
$\mu$ decay  
when $q^2 \rightarrow 0$ \cite{degrassi}. Following the general
argument sketched  
in Appendix A, we get the relation:
\begin{equation}
\alpha_2(q)^{-1}= \frac{\pi}{\sqrt{2} G_{\mu} m^2_W} 
+ \Pi^L_W(0)- \Pi_W^L(m_W)
+\Pi^T_W(q)-\Pi_W^T(m_W)
+ 2\Gamma^{U}_W(q)-\delta_{W}(0)  \,.
\label{alpha2}
\end{equation}
All the functions \cite{sirlin} $\Pi^L_W$, $\Pi^T_W$, $\Gamma^U_W$
and $\delta_W$  
(complete vertex and box contribution to the $\mu$--decay) are $bare$
functions. The divergences cancel out in the differences, the
same than in Eq. (\ref{eff}). 

Setting $q^2=0$ in Eq. \ref{alpha2} we get the value of $\alpha_2(0)$. 
Notice that $\alpha_2(0)$ can be expressed as,
\begin{equation}
\frac{G_{\mu}}{\sqrt{2}}= \frac{\pi \alpha_2(0)}{2 m^2_W(0)}
(1-\delta^{PD}_{W} (0))\,.
\end{equation}
Apart from the ``process--dependent'' term (which can be included in a
redefinition of the $G_{\mu}$ as a universal Fermi constant), we see
that the value $\alpha_2(0)$ 
can be related more directly to an experimental quantity
\cite{kuroda}, in an analogous way to the extraction of $\alpha_e(0)$
from Compton scattering. 
Therefore we have a close expression for both \alz{e}\ and \alz{2}\ in terms
of the respective values at zero momentum:
\begin{equation}
\alpha^{-1}_i(\mz)=\alpha^{-1}_i(0)\left[ 1+\Delta \alpha_i(\mz) \right]\,,
\label{alphaiz}
\end{equation}

If we want to compute instead the \ms--couplings, we get (see Eq.(\ref{rel})):
\begin{eqnarray}
\widehat{\alpha}_i^{-1}(\mz)&=&
\alpha_i^{-1}(0) -\left(\widehat{\Pi}^T_{i}(0)+2
\widehat{\Gamma}^U_i(0)\right)\\ 
&=& \alpha^{-1}_i(0)\left[ 1+\Delta \widehat{\alpha}_i(0) \right]\,.
\label{alphaizms}
\end{eqnarray}
Complete susy thresholds effects are included through their contribution
to the functions $\widehat{\Pi}_i$ and $\widehat{\Gamma}_i$, where the
divergent term has been subtracted, and the renormalization scale has
set to \mz.

The explicit values of $\alpha_e(\mz)^{-1}$ and $\alpha_2(\mz)^{-1}$
depend on the susy spectrum 
considered, but as a general result they are typically larger than 
the corresponding values of $\widehat{\alpha}_i(\mz)^{-1}\mid_{SM}$. 
On the following, we will use the notation $\widehat{\alpha}_i$ to
denote \ms\ couplings, and the 
subscript $SM$ when we do not include susy degrees of freedom, only
the standard ones. We will use the simplest parameterization of the susy
spectrum, assuming universal soft susy breaking terms at the GUT
scale, and neglecting the mixing between charginos and higgsinos, and
stops left and right. With these simplifications, we need only six
mass parameters for the matter spectrum: 
$m_{1/2}$ for the gaugino masses ($m_i\approx c_i
m_{1/2}$, $c_{\tilde{g}}=3$, $c_{\tilde{w}}=1$); $m_{1/2}$ and $m_0$
(common scalar mass at the GUT scale) for sleptons and squarks
($m_i=\sqrt{c_i m_{1/2}^2+m_0^2}$, $c_{\tilde{l}}=0.5$,
$c_{\tilde{r}}=0.15$, $c_{\tilde{q}}=7$); a common mass parameter,
$m_{\tilde{h}}$, for the higgsinos; a common mass, $m_H$, for the
heaviest Higgses, and $m_h$ for the Standard lightest Higgs; and also
$m_t$ for the top mass. We will take the susy mass parameters to be
not larger than $1\,TeV$ (naturalness bound). From the experimental
searches, we have the lower bounds: $m_{1/2} \geq 65\, GeV$ and $m_{h}
\geq 60\, GeV$. The susy parameters $m_{\tilde{h}}$ and $m_H$ will be 
taken at least of $O(\mz)$. For the top mass, we have the 
data: $m_t=176 \pm 8\pm 10$ (CDF) and $m_t=199 \pm 20^{+19}_{-21}$
(D0) \cite{top}. For numerical calculations we will allow  $m_t=200\,GeV$. 

With these constraints, and also assuming $m_0=m_{1/2}$ and 
$m_{\tilde{h}}=m_H$, we can write the numerical values of 
the effective couplings at \mz\ like:
\begin{eqnarray}
\alpha_e(\mz)^{-1}&=& 129.08 
+1.0\times 10^{-2} \Delta_t(176\, GeV) \nonumber \\
& &+3.9\times 10^{-4} \Delta_{1/2}(1\, TeV)
+1.8\times 10^{-4} \Delta_H(1\,TeV) \pm 0.3 \,,
\\
\alpha_2(\mz)^{-1}&=& 30.087  
+1.5\times 10^{-2} \Delta_t(176\,GeV)
+2.3\times 10^{-2} \Delta_h(60\, GeV) \nonumber \\
& &+2.4\times 10^{-4} \Delta_{1/2}(1\, TeV)
+1.7\times 10^{-4} \Delta_H(1\,TeV) \pm 0.09 \,,
\end{eqnarray}
with,
\begin{equation}
\Delta_i(m_0)= 1 - \left(\frac{m_0}{m_i}\right)^2\,.
\end{equation}
For arbitrary susy masses, these values are
typically 1\% larger than those of $\widehat{\alpha}_i^{-1}(\mz)\mid_{SM}$
\cite{pdg}:
\begin{equation}
\widehat{\alpha}_e^{-1}=127.9\pm0.3\,, \;\;\; 
\widehat{\alpha}_2^{-1}=29.66\pm0.09 \,.
\end{equation}

Due to the decoupling of the massive degrees of freedom in the effective
couplings, the values quoted above for susy masses $O(1\,TeV)$ are practically
the same as those we would get for the SM effective
couplings. Therefore, the initial 
``increasing'' of the effective couplings with respect to the
\ms--couplings at \mz\, is not due to the susy
contributions, or in general to any massive contribution, but due to
those coming from light quarks and leptons, which can be considered as
massless at the \mz\ scale. For example, the contribution of a ``light''
fermion ($m_f$ much less than \mz) to the effective coupling would be:
\begin{equation}
\Delta \alpha_i^{(f)}(\mz)= \alpha_i(0)\frac{2}{3 \pi} T_i^f \left(
\frac{5}{6}+\log \frac{m_f}{\mz} \right) \,,
\label{pif}
\end{equation}
whereas the \ms\ contribution is:
\begin{equation}
\Delta\widehat{\alpha}_i(0)^{(f)}=\alpha_i(0)\frac{2}{3 \pi}
T_i^f\log \frac{m_f}{\mz} \,. 
\end{equation}
In computing both \alz{i}\ and \alzms{i}, we run the couplings
from zero momentum to \mz, so that all the light thresholds are crossed. But
for \alzms{i}\ one uses no more than the step--function approximation,
while for the effective couplings we use a smother function
\cite{mdsp,clavelli}. That is the origin of the constant factor in
Eq. (\ref{pif}), and 
when we sum over all the ``light'' fermions, the main reason of the
difference between \alz{i}\ and \alzms{i} \cite{goldman}. 

The value of the QCD coupling will be derived
imposing unification within MSSM. However, the value
extracted from the experiment is the \ms\ couplings valid for the
SM. In order to compare, we have to eliminate the contribution due to
the susy degrees of freedom \cite{bagger}, and to  
change from ``effective'' to``\ms'' when required, that is,
\begin{eqnarray}
\widehat{\alpha}_3(\mz)^{-1}\mid_{SM}
 &=& \widehat{\alpha}_3^{-1}(\mz) +\left(\widehat{\Pi}^T_g(m_Z)+ 
\widehat{\delta}_3(\mz)\right)\mid_{susy}+\frac{1}{4 \pi}
\label{al3ms}\\
& = &\alpha_3(\mz)^{-1}
-\left(\widehat{\Pi}^T_g(\mz)+ 2 \widehat{\Gamma}^U_3(\mz)\right)\mid_{MSSM}
+\left(\widehat{\Pi}^T_g(m_Z)+\widehat{\delta}_3(\mz)\right)\mid_{susy}+
\frac{1}{4\pi}    
\label{al3eff} \,,
\end{eqnarray}
where the factor ``1/$4\pi$'' is due to the change in the
regularization procedure (from dimensional reduction to dimensional
regularization) when working with the SM. 
In the next section we will show that in fact Eqs. (\ref{al3ms}) and
(\ref{al3eff}) yield the same result. 

 Like for the other two couplings, the main difference between the
effective coupling \alz{3}\ and $\alzms{3}\mid_{SM}$ is due to the
change in the renormalization scheme, that is, to the contribution of the
massless degrees of freedom. We have not only the light
quarks, but also the gluon contribution \cite{mar},
\begin{equation}
\left(\widehat{\Pi}^T_g(\mz)+ 2
\widehat{\Gamma}^U_3(\mz)\right)\mid_{gluon}=-\frac{C_2(SU(3))}{4
\pi}\left(-\frac{11}{3}
\ln\frac{\mzz}{\mu^2}+\frac{157}{36} \right)\,. 
\end{equation}
Both together make the value of the effective QCD coupling roughly an
8\% larger than $\widehat{\alpha}_3(\mz)\mid_{SM}=0.117\pm0.005$ \cite{pdg}.

 The value of $\widehat{\alpha}_3(\mz)\mid_{SM}$ quoted above is the
average of a set of values coming from different experiments. 
Contrary to the situation with $\alpha_e$ or $\alpha_2$, in QCD we do
not have a natural experimental process to extract the value of
$\alpha_3$. The difference is obvious, because there is no limit to
zero--momentum transferred in QCD. We have a collection of
physical observables, which can be used to define ``effective
couplings'' taking into account the entire radiative correction into its
definition, one for each process, and can be related among them
\cite{effqcd}; for example, 
\begin{equation}
R(Q)\equiv \frac{11}{3}\left( 1+ \frac{\alpha_R(Q)}{\pi} \right)\,,
\end{equation}
being $R(Q)$ the total hadronic cross section in $e^+ e^-$
annihilation. 
 On the other hand, one can select a particular renormalization
scheme, say \ms--coupling, and express each observable like a series
expansion in this parameter, like \cite{alphar},
\begin{equation}
R(Q)=\frac{11}{3} \left( 1+ \frac{\almsi{3}}{\pi}+1.4092
\left(\frac{\almsi{3}}{\pi} \right)^2 + ... \right)\,.
\end{equation}
 Infinite series will return exactly the
renormalization--scheme invariant experimental quantities. But in
practice, we have available only finite order series, which can lead
to different theoretical predictions depending on the expansion
parameter chosen. In that sense, not all the couplings will be
reliable for all the processes.  Moreover, the effective couplings
like $\alpha_R$ are process--dependent by definition. The choice of
the best expansion parameter, and how to set its scale
\cite{scaleqcd}, is a major point of discussion in 
making theoretical predictions for QCD. The values obtained using
different convention may be quite different. Using the
renormalization group equations (RGE) to get $\alpha_R(\mz)$ from
$\alpha_R(31.6\,GeV)=0.165 \pm 0.016$ 
\cite{exper}, 
one gets a value of $O(10\%)$ larger than $\almsi{3}(\mz)$.  

Another example is provided by the ``momentum--scale'' subtraction QCD
coupling \cite{celmaster}, which is related to the \ms\ coupling by,
\begin{equation}
\alpha_3^{mom}=\almsi{3} \left( 1+ A(n_f) \frac{\almsi{3}}{\pi} +
... \right)\,,\;\;\;\; A(5)=1.9776\,. 
\label{mom}
\end{equation}
Again, we would obtain $\alpha_3^{mom}$ of roughly a 7\%
larger than $\widehat{\alpha}_3$. The definition of $\alpha_3^{mom}$
is gauge and process 
dependent, that is, depends on the vertex chosen to set the
renormalization constant (trigluon vertex, quark--gluon,...). In
Eq. (\ref{mom}) $\alpha_3^{mom}$ is given in the Landau gauge and for
the trigluon vertex. Other possible choices do not change appreciable
the numerical factor $A(5)$.

The problem of gauge dependence of the effective charges
\cite{coquer}, in the sense of explicit presence of 
the gauge parameter in the constant contribution, also afflicts the
definition we use. In order to minimize their effect in the evolution
with the scale, we work in the Landau gauge, which is a fixed point of
the RGE for the gauge parameter. This problem can be solved including
the appropriate box corrections. However, these are process--dependent
corrections, as those coming from the vertex. In order to have some
kind of universal QCD coupling we would need to set some convention to
define the process independent contribution.

In this line it works the effective QCD coupling defined by the
interaction potential between two infinitely massive quarks
\cite{qcdpot}, in the same spirit than the pure QED effective coupling,
\begin{equation}
V(Q)\equiv - \frac{4 \pi C_F \alpha_v(Q)}{Q^2}\,.
\end{equation}
Threshold effects are associated with the radiative corrections to the
propagator of the exchanged gluon, rather than the vertex or box
corrections. Therefore, they are universal, and vertex and boxes are
only intended to ensure the gauge independence of $\alpha_v$. In
principle, $\alpha_v$  (and its extension to the supersymmetric
theory) would provide a good scheme to deal with thresholds. However,
in order to study unification (our main motivation), we should extend
this scheme, or any other, to define $\alpha_e$ and $\alpha_2$, with
the additional 
complication that these couplings are related to a broken gauge
symmetry above \mz. Because of that, we have at first set the
renormalization scheme for the effective $\alpha_e$ and $\alpha_2$,
and extended it to $\alpha_3$ afterwards. For the broken theory, we
use the fact that the ``universal vertex correction'' is related to
the longitudinal term of the mixed vacuum polarization tensor for the
neutral bosons \cite{kennedy}. 

If we  correctly set the relation between different schemes, it does
not matter which renormalization scheme we consider, as far as we know
what are the physical effects included, say threshold effects. 
This is the last step in order to compare
the theoretical predictions with the available experimental data.

\section{ Unification with heavy thresholds} 
 A real unification picture of
the gauge couplings implies not only to get a common value at some
point in the high energy sector, but a common evolution beyond some
scale up to the Planck scale, which can be identified with the value
and evolution of the gauge coupling associated with the unification
group. This can be obtained only through the modification of the
running of the couplings due to new (heavy) degrees of freedom
coming from  the unification group \cite{zichichi}. 
We will fix, when needed, the unification group to be $SU(5)$. 
The same problem of accuracy of crossing the heavy thresholds will appear 
again in the high energy region. 

The heavy mass spectrum of \su5\ is given as usual in terms of 3 mass
parameters: \mv, for the heavy gauge boson masses, \mphi\ for the
color triplet Higgs, and \msig\ for the scalars in the adjoint. After
including the contribution of these new particles in the running of
the effective couplings, we will get real unification above some scale
{\it larger than the largest heavy mass parameter}. Therefore, we will fix
the unification condition for the gauge couplings at the Planck scale,
\mp, that is,
\begin{equation}
\alpha_1^{-1}(\mp)=\alpha_2^{-1}(\mp)=\alpha_3^{-1}(\mp)\,.
\label{unieff}
\end{equation}

The expressions for the effective gauge couplings, including also the
heavy degrees of freedom coming from \su5\, are given in Appendix B,
at 1--loop and 2--loops order. Thresholds at 2--loop order are treated
in an approximate way. In fact, we neglect those of the light massive
degrees of freedom in the 2--loop coefficients. The inclusion of more
detailed 2--loop threshold functions would introduce only a
modification less than 1\% in the running of the couplings. However,
we can not forget about the contribution of the heaviest masses at
2--loops, as we need to get at the end of the energy scale the same
evolution for the three couplings. As we know that these degrees of
freedom are completely decoupled well below their mass scale, we use
for these masses the step--function approximation in the 2--loop
coefficients.

If we work instead with the couplings \alms{i}, susy thresholds are
included in the initial values at \mz, but there is no indication
about how to cross the heavy ones. A consistent approach would be to
integrate out these heavy degrees of freedom from the complete action
$S[G]$ \cite{weinberg}, and in this way one gets the unification
condition: 
\begin{equation}
\widehat{\alpha}_i(\mu)^{-1}= \widehat{\alpha}_G^{-1} +
\lambda_i(\mu)\,,
\label{unims}
\end{equation}
where $\mu$ is a scale much larger than all light masses, and much
smaller than the heavy masses\footnote{``much larger, much smaller''
means at least two orders of magnitude of difference.}. The function
$\lambda_i(\mu)$ is given at 1--loop order by:
\begin{equation}
\lambda_i(\mu)=-\frac{1}{2\pi} \sum_{j=Heavy} b^{(j)}_i \ln
\frac{M_j}{\mu}\,.
\end{equation}

 It is straightforward to show now that both equations (\ref{unieff})
and (\ref{unims}) are completely equivalent, and thus we will obtain
the same predictions for \alzms{3}, that is, for $\alzms{3}\mid_{SM}$,
with both approaches.

First, we compute the value of $\alpha_i^{-1}(\mu)$ from
$\alpha_i^{-1}(\mp)=\alpha_G^{-1}(\mp)$, that is,
\begin{equation}
\alpha_i^{-1}(\mu)=\alpha_G^{-1}(\mp)+F_i(\mu)-F_i(\mp)\,,
\label{almu}
\end{equation}
where in order to simplify the notation, we have defined:
\begin{equation}
F_i(q)= \Pi^T_i(q)+2 \Gamma^U_i(q)\,.
\end{equation}
 Using the relation between effective couplings and \ms\ couplings, we
get:
\begin{eqnarray}
\alpha_i^{-1}(\mu)-\widehat{F}_i^{(l)}(\mu) & = & \alpha_G^{-1}(\mp) -
\widehat{F}_i(\mp)+\widehat{F}_i^{(H)}(\mu)\,, \\ 
\widehat{\alpha}_i^{-1}(\mu) &=& \widehat{\alpha}_G^{-1}(\mu)+
\widehat{F}_i^{(H)}(\mu)\,,
\end{eqnarray}
where $\widehat{F}_i$ means that the divergence $2/\epsilon$ has been
subtracted, and now the renormalization scale has been set to $\mu$; the
subscript ``H'' indicates only heavy degrees of freedom, while ``l''
refers to the light particles. The function $\widehat{F}_i^{(H)}(\mu)$
is that we obtain when the heavy degrees of freedom are integrated
out, i.e., that we call above $\lambda_i(\mu)$.

In this derivation, no reference is made to the order of perturbation
theory we were working, and therefore the equivalence between both
approaches is maintained at $any$ order of perturbation
theory. However, an additional result obtained, working with
\ms\ couplings, is that the function $\lambda_i(\mu)$, which includes the
information about the heavy spectrum, {\it should be computed at the
same order of perturbation theory than the couplings \alms{i}}.

It is commonly assumed that if we run the couplings at 2--loop order 
we need only $\lambda_i(\mu)$ at 1--loop, i.e., 2--loops heavy
thresholds correction would be a higher order 
correction, and thus, negligible. That argument relies on the fact
that these correction are $O(\widehat{\alpha}_G)$, and therefore in
principle negligible in the R.H.S. of Eq. (\ref{unims}),
\begin{equation}
\widehat{\alpha}_i^{-1}(\mu)=\widehat{\alpha}_G^{-1}(\mu) 
-\frac{1}{2\pi} \sum_{j=Heavy} b^{(j)}_i \ln \frac{M_j}{\mu} +
O(\widehat{\alpha}_G)\,.
\end{equation}
Nevertheless, when $\widehat{\alpha}_i^{-1}(\mu)$ is computed at
2--loop order these are the kind of corrections,
$O(\widehat{\alpha}_j(\mu))$, which are taken into account in the
L.H.S. From this point of view, there is no any reason to neglect them
in the R.H.S.

That argument would also imply that working at 1--loop order it is not
needed any heavy threshold correction in the unification condition,
that is,
\begin{equation}
\widehat{\alpha}_i^{-1}(M_X)= \widehat{\alpha}_G^{-1}(M_X) \,,
\label{1uni}
\end{equation}
where $M_X$ would be some point in the high energy region. But that is
no more than the unification condition when we consider a
complete degenerate heavy spectrum, being $M_X$ the heavy mass
scale. Heavy threshold corrections depend 
on the degree of degeneracy of the spectrum, independently of the
order of perturbation theory we work with. Eq. (\ref{1uni}) can not be
considered as the unification condition for a more general
heavy spectrum, even at 1--loop. Moreover, that would not be
compatible with the 
picture obtained with the effective couplings. And if unification is a
physical process, it should be independent of the renormalization
scheme we use to study it.

In a more general case, notice that when we include only 1--loop heavy
thresholds 
corrections in the running of 2--loop \ms\ couplings, we end up with a
dependence on the scale ``$\mu$'' to which the unification
condition is imposed. On one hand, when we integrate out the heavy
fields from the action, that scale has to be much smaller than the
heavy masses if we want to keep only the dominant logarithmic
contributions in $\lambda_i(\mu)$. On the other hand, one can prefer
$\mu \approx M_j$ in order to avoid large corrections to the relation
between the couplings. But all this arbitrariness in $\mu$ disappears when
we work with $\lambda(\mu)$ at the same order as that the couplings, and
we do not have to worry about any specific choice. 
The RGE guarantees that Eq. (\ref{unims}) is {\it scale invariant}, when all
the terms involved are computed at the same order in perturbation 
theory. Taking the derivate respect to $\ln \mu$ we get,
\begin{eqnarray}
\frac{d \lambda_i(\mu)^{(1-loop)}}{d \ln \mu} &=& \frac{1}{2
\pi} (b_G^{(1)}-b_i) = \frac{1}{2 \pi} \sum_{k=heavy} b_i^{(k)} 
\theta (M_k-\mu)\,, \\
\frac{d \lambda_i(\mu)^{(2-loop)}}{d \ln \mu} &=& \frac{1}{8
\pi^2}(b_G^{(2)}\widehat{\alpha}_G(\mu)
-\sum_j b_{ij} \widehat{\alpha}_j(\mu))
\approx \frac{1}{8 \pi^2} 
\sum_{k=Heavy} \sum_{j} b_{ij}^{(k)}\widehat{\alpha}_G(\mu)\theta (M_k-\mu)\,, 
\end{eqnarray}
where $b_G^{(1)}$, $b_G^{(2)}$ are the 1--loop and 2--loops
coefficients for the unification group $G$.   
The uncertainty introduced
when neglecting $\lambda_i(\mu)^{(2-loop)}$ will depend on the nature
of the unification group, and mainly on the degree of degeneracy of
the heavy spectrum. For a nearly degenerate spectrum, the
choice $\mu \approx M_k$ will minimize the contribution of
$\lambda^{(2-loop)}$ (and also $\lambda^{(1-loop)})$. In the case of
\su5, with only 3 relevant heavy mass parameters, even if we allow a
different of 2 to 3 orders of magnitude among them the correction
$\lambda_i^{(2-loop)}$ would not  change the prediction of \almsi{3}\
by more than $O(1\%)$. However, this may be not the case once the
heavy spectrum is enlarged. For example, in the Missing Doublet $SU(5)$ Model
\cite{mdm} the
scalar that breaks the symmetry down to the MSSM is contained in the
75 representation of $SU(5)$, instead of using the 24--Higgs. However, the 75
fields are not degenerate in mass, and the ratios of their masses will
contribute to the prediction of \alms{3} \cite{yamada,mar2}. These constant
factors can 
enhance the 1--loop prediction by a factor of $O(12\%)$. Due to the
presence of large $b_{ij}^{(k)}$ coefficients, this sector will also
makes an important contribution to order 2--loops, even an increasing
of $O(10\%)$.

 To avoid such uncertainties at the 2--loop level,
we adopt the same kind of approach to treat 2--loop heavy mass
corrections than with the effective couplings. That is, we include
their contribution in the $b_{ij}$ coefficients using the
step--function approximation, and demanding unification at a scale
larger than the heaviest mass, say the Planck scale.

 The value of $M_V$ and the unification
gauge coupling are derived together with $\alpha_3(m_Z)$ from the
unification condition. The value of \mphi\ is bounded by the limits on
proton decay via dimension-five operators \cite{hisano}. The minimum
allowed value of \mphi\ will depend on the masses of gauginos, squarks
and sleptons, decreasing 
with the ratio $\xi_0=(m_0/m_{1/2})^2$. On the other hand, 
the value of $\alpha_3(m_Z)$  decreases
when the susy masses are raised, and increases with \mphi. The minimum
value for the QCD coupling is obtained for squark and higgsino masses
of $1\,TeV$ 
(naturalness bound) and $m_{1/2}\simeq 70\,GeV$. In Table I we have
given the minimum value of $\alpha_3(m_Z)\mid_{SM}$ obtained with both the
effective couplings and \ms, for different values of \msig. At the 2--loop
order we have a mild dependence on this variable in the value of
\alzi{3}. Nevertheless, \msig\ affects mainly the prediction of \mv. We
can get a large value of $M_V$ (and therefore of the unification
scale) just diminishing enough that of \msig.

We can see from Table I that Susy $SU(5)$ unification requires 
$\alzi{3} \ge 0.127 $ with a susy spectrum not larger than $1\,TeV$
\cite{bagger}. Notice that the value of $M_X$ quoted
for a degenerate heavy spectrum would not be compatible with the
constraints on proton decay. To get a larger $M_X$ we have to reduce
the gaugino mass, and therefore we would increase the value of
\alzi{3}. Nevertheless, we do not aim to remark these numerical values
as in our study several effects suitable of changing them were not
included. In the first place, we have not taken into account the Yukawa
contribution to the 2--loop running of the Yukawa couplings, just for
the sake of simplicity. This correction is not expected to lower
\alzi{3} more than an 1\%. 

The second correction not included is 
that due to non renormalizable operators coming from quantum
gravitational effects \cite{nro}, which begin to be relevant as we
approach the Planck scale. Although these operators are 
suppressed by a factor \mx/\mp\, their unknown strength may introduce
a large correction which can have either sign. At present, this unknown
factor would enlarge the allowed range for \alzi{3}\ to be compatible
with any experimental value. On the other hand,
 a more precise measurement of the QCD coupling together with the
observation of proton decay (which would give the value of \mv\ or
\mphi) can constrain the strength of the gravitational effects
\cite{nathar}.  

\section{ Concluding remarks}
The precision reached in the experimental extraction of $\alpha_e$ and
$\alpha_2$ has promoted during the last years the study of
supersymmetric unification beyond the lowest order approximation. This
leads to the inclusion of 2--loop effects in the running of the gauge
couplings, together with a proper treatment of light and heavy
threshold effects. Here we have focussed on the treatment of light
thresholds beyond the leading log approximation, and the related topic
of renormalization scheme dependence. 

 Physical processes are renormalization scheme independent, but it is
not so for the gauge coupling parameters. The latter are extracted
from the physical quantities using the \ms\ scheme and assuming the
SM. To study unification in the MSSM we can choose different 
schemes to set the running of the couplings. Using \ms\ scheme complete light
thresholds are included in the initial value of the couplings at \mz,
but not in their evolution. Using effective couplings, the values at
\mz\ are also modified by the presence of massive degrees of freedom,
but contributions from large masses are decoupled at \mz. When running
the couplings, the Mass Dependent RGE gives us the correct crossing of
the thresholds. We have explicitly shown that both schemes gives the
same prediction for the QCD gauge coupling, once the conversion to the
\ms\ scheme and the SM is done (see Table I). These values are obtained
at 2--loop order including also heavy threshold contributions, which
are important when the heavy spectrum is non degenerate (as can be
expected in realistic models). Renormalization group arguments show
that heavy thresholds have to be included at the same order of
perturbation theory we run the couplings. In minimal \su5\ one does
not expect 2--loop heavy thresholds to be large; however, this might
not be the case for other unification models with more heavy degrees
of freedom.  

 Working in MSSM with \ms\ has the obvious advantage that this is the
renormalization scheme used to give the experimental data in the
SM. The conversion only requires the suppression of the new degrees of
freedom from the values of the gauge couplings. It has the
disadvantage that there is no information in this scheme about how to
treat new thresholds. The initial values $\almsi{i}(\mz)$ for the MSSM
will not be valid if we allow for example the presence of extra matter
\cite{babu} 
at a scale larger than \mz\ (but below the unification scale). The
values of $\alzmsi{i}$ including thresholds would not be a good
indication of  
the strength of the interaction at that scale, unless we impose
decoupling. For each model, we have to readjust both  the
$\beta$--functions and the initial conditions. The situation is
different for the effective couplings. The matter content
at the scale \mz\ (masses near this scale) fixes the values of the
couplings. The introduction of heavier degrees of freedom is done
trough their $\beta$--functions when the couplings evolve with the
scale, which takes into account a smooth threshold crossing.

\acknowledgments 
 I would like to thank B. Brahmachari, J. C. Pati, K. S. Babu,
S. Brodsky and J. P\'erez--Mercader for very enlightening discussions. 

\widetext
\begin{table}
\caption{
Minimum value of $\alpha_3(m_Z)\mid_{SM}$ obtained with both the
effective couplings and \ms, for different values of \msig\ and
$\mphi=10^{16.73}$. The latter is the lower value consistent with
proton decay when the 
susy masses are $m_{\tilde{q}}=m_{\tilde{h}}=1\,TeV$, $m_{1/2}=71\,
GeV$ ($\xi_0\simeq 190$). We also quote the value of \alzi{3}\ for the
MSSM. In the case of degenerate heavy spectrum, $m_{1/2}=377\,GeV$
($\xi_0=0.1$) .
}
 \begin{tabular}{c|cc||ccc} 
 &&\ms\ couplings & Effective  couplings && \\
\tableline
\msig\ & $\alms{3}\mid_{SM}$ &\mv\ 
&$\alzi{3}\mid_{SM}$ & \mv & $\alzi{3}\mid_{MSSM}$ \\
\tableline
\begin{tabular}{c}                     
Degenerate  \\ Heavy spectrum          
\end{tabular}  &  0.1220 & $10^{16.10}$ &  0.1232 & $10^{16.14}$ & 0.1345 \\
$10^{14}$      &  0.1264 & $10^{17.70}$ &  0.1269 & $10^{17.76}$ & 0.1391 \\
$10^{15}$      &  0.1268 & $10^{17.15}$ &  0.1273 & $10^{17.21}$ & 0.1397 \\
$10^{16}$      &  0.1272 & $10^{16.61}$ &  0.1277 & $10^{16.66}$ & 0.1401 \\
 \end{tabular}
\end{table}
\narrowtext

\appendix
\section{}

 In this Appendix we give the general expression for the relation
between experimental quantities and effective couplings. We want to
keep the discussion as general as possible, so we do not make use of
any particular experimental value.
 Let us assume instead that we have available the
experimental value for the transition amplitude, $A_{exp}$, obtained
from the scattering process at the scale $q_0$, mediated by the gauge
boson associated with the coupling $\alpha_i$. Once the radiative
corrections are taken into account, we can write down the expression
for $A_{exp}(q_0)$ like:
\begin{equation}
A_{exp}(q_0)= \frac{\alpha_{i0}}{q_0^2-m_{i0}^2}
\left\{ 1 - \alpha_{i0}\frac{\Pi_i(q_0)}{q^2_0-m_{i0}^2}- \alpha_{i0} 2 \Gamma_i(q_0) - (q_0^2-m_{i0}^2) \alpha_{i0} B_i(q_0) \right\} \,,
\label{Aexp}
\end{equation}
where $m_{i0}$ and $\alpha_{i0}$ are the bare mass and coupling, and
$\Pi_i$, $\Gamma_i$ and $B_i$ are the bare vacuum polarization, 
vertex and box contributions respectively (defined without the
factor $\alpha_{i0}$, that has been written explicitly). 
For the gauge vacuum polarization tensor, we follow the convention:
\begin{equation}
\Pi_i^{\mu\nu}(q)=(g^{\mu \nu} q^2-q^{\mu} q^{\nu}) \Pi_i^T(q)+
m_{i0}^2 g^{\mu\nu} \Pi_i^L(q)
\end{equation}
where ``T'' and ``L'' have the usual meaning of transverse 
and longitudinal terms.

The L.H.S. of Eq. (\ref{Aexp}) is both gauge invariant and finite. The
bare mass can be replace in terms of the physical mass, $m_i$, through
the equation:
\begin{equation}
m_i^2=m_{i0}^2-\alpha_{i0} \Pi_i(m_i)\,.
\end{equation}
And we obtain:
\begin{equation}
\frac{A_{exp}^{-1}(q_0)}{q^2_0-m^2_i} =
\alpha_{i0}^{-1}+\frac{\Pi_i(q_0)-\Pi(m_i)}{q^2_0-m_i^2}+2 \Gamma_i(q_0)
+(q_0^2-m_i^2) B_i(q^0)\,.
\label{Aexpi}
\end{equation}
The remaining divergences of the L.H.S. of Eq.(\ref{Aexpi}) 
will cancel out when
we replace the bare coupling by the renormalized coupling, whatever
the renormalization scheme we use. Instead of working this way, let us
use Eq. (\ref{Aexpi}) like a definition for the bare
coupling. Therefore, this can be used to get, for example, the
renormalized \ms\--coupling, replacing back $\alpha_{i0}$ in the
definition
\begin{equation}
\widehat{\alpha}_{i}^{-1}(\mu)=Z_i \alpha_{i0}^{-1} \,,
\label{defms}
\end{equation}
where $Z_i$ is the corresponding product of renormalization constants,
and $\mu$ is the renormalization scale. Or we can get the effective
coupling, $\alpha_i(q)$, in a similar way:
\begin{eqnarray}
\alpha_i^{-1}(q)&=&\alpha_{i0}^{-1}
+\left( \Pi^T_i(q)+2 \Gamma_i^{U}(q) \right) \label{defeff} \\ 
&= &
\frac{A_{exp}^{-1}(q_0)}{q^2_0-m^2_i}-\frac{\Pi_i(q_0)-\Pi(m_i)}{q^2_0-m_i^2}
+ \Pi^T_i(q)-\delta_i(q_0)+2 \Gamma_i^U(q) \,.  
\label{ali}
\end{eqnarray}
In the last line we
have arranged the vertex and box contributions in the function
$\delta_i$. Notice that the function $\Gamma_i^U$ is not the complete
vertex that appeared in Eq. (\ref{Aexp}), but the process--independent
(universal) part of this function.

A final remark about the functions $\Pi_i^T$ and $\Gamma_i^U$. Those
functions involved in the definition of the effective couplings
(Eq. (\ref{defeff})), are defined in Euclidean space--time, so 
that in some sense we are working with ``Euclidean'' effective
couplings. We make this choice instead of keeping the momentum in
Minkowski space--time because we were interested in dealing with
continuous differentiable functions when crossing the thresholds. On
the other hand, the functions involved in Eqs. (\ref{Aexp}) and
(\ref{Aexpi}) are defined using Minkowski momentum, and that produces
both kind of behaviors to be mixed in the relation (\ref{ali}). This
is perfectly consistent. However, if we had defined ``Minkowski''
effective couplings, we would have a more direct relation between
those and physical quantities. For example, setting $q_0=q$ and
$m_i=0$ (this would be the case for the QCD coupling) in
Eq. (\ref{ali}), we would get:
\begin{equation}
\alpha_i^{-1}(q)= \frac{A^{-1}_{exp}(q)}{q^2}- \delta_i^{PD}(q)\,,
\end{equation}
the last term being the process dependent contribution of vertex and
boxes to the physical amplitude. Neglecting this term, we would have a
direct measure of the effective coupling. With the Euclidean coupling
we get instead,
\begin{equation}
\alpha_i^{-1}(q)\mid_{Mink.}= \frac{A^{-1}_{exp}(q)}{q^2}- \delta_i^{PD}(q)
-(\Pi^T_i(q)\mid_{Mink}-\Pi^T_i(q)\mid_{Eucl})\,.
\end{equation}
The last difference is not negligible near the threshold of the
massive particles . A ``quasi'' direct measurement of the Euclidean
coupling is obtained only for scales $q$ that are far enough of any
threshold (below or beyond). Nevertheless, even if the Euclidean
effective couplings are not so nicely related to the physical
quantities as the Minkowski couplings, they both share the same kind of
behavior with respect to the very light degrees of freedom, and most
important, with respect to the very heavy degrees of freedom
(decoupling).

In order to keep a simple notation we have not distinguish throughout
the paper when is used the Euclidean momentum or the Minkowski
momentum. However, these can be easily identified from the precedent
discussion.

To end this appendix, we write also the relation between
$\widehat{\alpha}_i^{-1}$ and $\alpha_i^{-1}$, given by:
\begin{equation}
\alpha_i^{-1}(q)=\widehat{\alpha}_i^{-1}(\mu)
+\left( \widehat{\Pi}^T_i(q)+2 \widehat{\Gamma}_i^{U}(q) \right) \,,
\label{rel}
\end{equation}
that can be traced easily
for example from the definition of effective coupling
Eq. (\ref{defeff}) and that of the \ms\ couplings Eq. (\ref{defms}). 
The symbol ``$\widehat{ }$ '' over the functions $\Pi_i$ and $\Gamma_i$
means that the divergent term has been subtracted, and $\mu$ is the
renormalization scale.
The functions $\widehat{\Pi}_i^T$ and $\widehat{\Gamma}^U_i$ behave as
$\ln M_i/\mu$ when $M_i/q \rightarrow 0$, and therefore decoupling is
not present in the \ms\ couplings.

\section{}

 Here we give the expression for the effective couplings including the
heavy degrees of freedom coming from \su5. Their general expressions
are given in Appendix A of Ref. \cite{mar}. The 1--loop \su5\
contributions are given by,
\begin{equation}
(4 \pi) \left( \Pi^T_i(q)+2 \Gamma_i^{U}(q) \right)^{(heavy)} =
-\overline{C}_i G^{gauge}(\mv) + \frac{\overline{C}_i}{2}
G^{chiral}(\mv) + \sum_{a=\Phi,\Sigma} b_i^{a} G^{chiral}(M_a)\,, 
\end{equation}
where,
\begin{eqnarray*}
\overline{C}_i&=&C_2(SU(5))-C_2(G_i)=(5,3,2) \,,\\
b_i^{\Phi}&=&(2/5,0,1)\,, \\ 
b_i^{\Phi}&=&(0,2,3)\,, 
\end{eqnarray*}
and, 
\begin{eqnarray}
G^{gauge}(\mv) &=&
7 \left(\frac{2}{\varepsilon}-\ln\frac{-q^2}{\mu^2}\right)
+\frac{13}{3} F_{\sg}(\mv,\mv) +3 F_{\sgam}(\mv)+ F_{\ss \sg}(0,\mv) 
-\frac{1}{3} F_{\ss}(0,0) \\
G^{chiral}(M_a) &=&
\left(\frac{2}{\varepsilon}-\ln\frac{-q^2}{\mu^2}\right) 
+\frac{1}{3} F_{\ss}(M_a,M_a) +\frac{2}{3} F_{\sf}(M_a,M_a) 
\end{eqnarray}
The functions $F_{i}(M_j,M_k)$ are defined in \cite{mar}.

At 2--loop order we have,
\begin{equation}
\alpha_i^{-1}(\mu)=\alpha_i^{-1}(\mu)\mid_{1-loop} - \frac{1}{8 \pi^2}
\int_{\mz}^{\mu} b_{ij}(\mu')\alpha_j(\mu') d\ln\mu'\,.
\end{equation}
 In a Mass Dependent renormalization scheme, the coefficient
$b_{ij}(\mu')$ depends on the ratio of the masses and the scale. We
neglect the contribution of light thresholds and approximate those of
the heavy degrees of freedom by a step--function,
$\theta_k=\theta_k(\mu-M_k)$. The $b_{ij}$ 
coefficients for the MSSM are given in Ref. \cite{2loopb}, and  the heavy
contribution for the matter content of \su5\ is given by,
\begin{eqnarray}
b_{11} &=& \frac{232}{3}\theta_V+\frac{167}{25}\theta_V
\theta_\Phi+\frac{2}{75}\theta_\Phi \nonumber\,, \\ 
b_{12} &=& 15 \theta_V \nonumber \,, \\
b_{13} &=& \frac{80}{3}\theta_v+\frac{32}{15} \theta_\Phi \nonumber \,, \\
b_{21} &=& 5 \theta_V \nonumber \,, \\
b_{22} &=& 24 \theta_V + 3 \theta_V \theta_\Phi 
+36 \theta_V\theta_\Sigma+ 24 \theta_\Sigma \nonumber \,,  \\
b_{23} &=& 16 \theta_V \nonumber \,, \\
b_{31} &=& \frac{10}{3}\theta_V + \frac{1}{15} \theta_\Phi \nonumber\,,  \\
b_{32} &=& 6 \theta_V \nonumber \,, \\
b_{33} &=& \frac{50}{3} \theta_V + \frac{31}{5}\theta_V \theta_\Phi
+36 \theta_V \theta_\Sigma+ \frac{34}{3} \theta_\Phi + 54 
\theta_\Sigma \nonumber \,.
\end{eqnarray}

\end{document}